\batchmode
\makeatletter
\def\input@path{{\string"C:/Users/roomate/Desktop/spintronic1/my paper/1/final/1/after prb reject/\string"/}}
\makeatother
\documentclass[twoside,english,prb,aps,manuscript]{revtex4-1}
\usepackage[T1]{fontenc}
\usepackage[utf8]{luainputenc}
\usepackage{units}
\usepackage{amsmath}
\usepackage{amssymb}
\usepackage{graphicx}
\usepackage{setspace}
\usepackage{esint}

\makeatletter
\batchmode

\def\input@path{{C:/Users/roomate/Desktop//}}

\usepackage{units}
\usepackage{setspace}
\usepackage{babel}

\makeatother

\usepackage{babel}
\begin{document}

\title{Spin Hall effect originated from fractal surface}

\author{I. Hajzadeh$^{1}$, S.M. Mohseni$^{1,*}$, S.M.S. Movahed$^{1}$,
G.R. Jafari$^{1}$}

\address{$^{1}$Department of Physics, Shahid Beheshti University, Velenjak,
Tehran 19839, Iran}
\begin{abstract}
Spin hall effect (SHE) in thin films is inherited by surface roughness.
Although roughness effect on SHE has been studied in thin films, but
roughness is not only parameter in rough surfaces. Our results show
that how other statistical parameters of rough surface play important
role in SHE. In this paper we investigate theoretically the effects
of correlated surface roughness in the SHE with self affine fractal
surface in non-heavy metallic thin films in the frame work of the
Born approximation. The surface roughness is described by the k-correlation
model and is characterized by the roughness exponent $H$ $\left(0\leq H\leq1\right)$,
the in plane correlation length $\xi$ and the $rms$ roughness amplitude
$\delta$. We show that the spin Hall angle can increase by one order
of magnitude when $H$ decreasing from $H=1$ to $H=0$. We also demonstrate
the SHE for surface roughness with distribution function of the Gaussian
profile is mainly contributed by the side jump scattering while for
that with a non-Gaussian profile, both side jump and skew scattering
are present. our achievements demonstrate the important role of roughness
texture profile for SHE in non-heavy metals.
\end{abstract}


\email{Corresponding author email address: m-mohseni@sbu.ac.ir, majidmohseni@gmail.com}

\maketitle

\section*{I. Introduction}

Spin Hall effect (SHE) and its inverse effect (ISHE) are a group of
phenomena emerge from spin-orbit coupling (SOC) in non-magnetic metals
and semiconductors. SHE converts electrical conductivity to transverse
spin Hall conductivity (SHC) in non-magnetic layer without using magnetic
field (ISHE acts in opposite way exactly) \cite{dyakonov1971current,hirsch1999spin,hoffmann2013spin,sinova2015spin,niimi2015reciprocal,maekawa2012spin}.
These two effects provide a possibility of detection and generation
of SHC in non-magnetic materials \cite{kato2004observation,wunderlich2005experimental,saitoh2006conversion,valenzuela2006direct,zhao2006coherence,miron2011perpendicular,liu2012spin,maekawa2012spin}
and have potential applications in spintronic devices, such as Spin-Hall
oscillators, SHE transistors, spin photodetectors, spin thermoelectric
converters, domain wall electronics and spin Hall magnetic memories,
etc \cite{miron2011perpendicular,uchida2010spin,jungwirth2012spin,nakayama2013spin,zhang2015role,chen2016spin,hoffmann2013spin,wunderlich2010spin,wunderlich2009spin,ando2010photoinduced,kirihara2012spin,liu2012spin}.

A key challenge for the promotion of such spintronic devices is to
attain efficient conversion between charge and spin currents. It has
been believed that heavy metals with strong spin–orbit interaction
are indispensable. This is largely restrictions for the selection
of materials for the practical application of the spintronic devices.
Light metals have been confirmed to exhibit negligible SHEs. Thus,
whether the SHEs can be enhanced in light metals is an important fundamental
and practical question to push forward the application of the spintronic
devices with a large selection of materials. In this study, we demonstrate
that light metals (Our analysis based on Cu thin film) becomes an
efficient converter between spin and charge currents through correlated
surface roughness and statistical parameters of rough surface. Recently,
the problem of surface roughness with uncorrelated surface profile
on the SHC of non-magnetic thin metallic films has been developed
by Zhou et al \cite{zhou2015spin}. But, an important question comes
with ``can uncorrelated roughness describe a real surface?''. AFM
images in thin film deposition show that thin film surfaces are correlated
and self-affine \cite{zhao2000characterization,Jannesar2014}. Also,
the height-height correlation function plays a significant role in
the limit $k_{F}\xi\gg1$, where $\xi$ is the in plane correlation
length for the surface roughness and $k_{F}$ is the Fermi wave vector
\cite{fishman1991influence}. To address these facts, we focus on
the correlated surface that have fractality effect. 

In this paper, to accomplish the argument about effects of self-affinity
in SHA in light metals, and effects of Gaussian and non-Gaussian of
distribution functions of surface roughness in SHC, we explore the
influence of the height-height correlation function on the SHA with
self-affine surface roughness. The surface roughness will be regarded
as effective impurities \cite{tevsanovic1986quantum,trivedi1988quantum}
and the power spectrum of isotropic rough surface will be described
by k-correlation model \cite{m,makse1996method,Zamani2012,Jamali2015}.
The self-affine fractal surface roughness is characterized in addition
to height fluctuations $\delta$, from flatness and $\xi$ by a local
fractal dimension $d_{f}=3-H$, where $H$ is the Hurts or roughness
exponent and $0\le H\le1$. Smaller value of $H$ and $\xi$ correspond
to the rougher surface. By introducing the spin-orbit interaction
associated with fractal effective impurity, we find that the main
interplay mechanism of the roughness effect on SHA occurs for $H$
and $\xi$, and SHA can increase by one order of magnitude, when the
roughness exponent varies from $H=1$ to $H=0$. Moreover, we find
that if the distribution function of surface roughness taken to be
Gaussian, the SJ contributes to the SHE and the skew scattering (SS)
does not have any contribution in SHC induced by surface roughness.
Contrarily, for that to be non-Gaussian, the SS contributes in SHC.

The rest of this paper is organized as follows. In Sec.II, we present
a theoretical description for a thin film with generalized correlated
surface roughness and the uncorrelated bulk impurities. In Sec.III,
using the transition probability and relaxation times, we obtain the
longitudinal conductivity and the SHC. Furthermore, we describe the
influence of Gaussian and non-Gaussian roughness distribution functions
on SHE. In Sec.IV, a typical model for a self-affine rough surface
characterized by power-law function for correlation are described.
We also discuss the routes to enhance the surface roughness induced
spin-orbit interaction due to the fractal surface scattering. Finally
some general summary and conclusions are presented in Sec.V.

\section*{II. Model}

We consider a metal thin film with correlated rough surface extended
in a plane with $\rho=(x,y)$ direction and confines in the $z$ direction
with variable thickness $d(\rho)$. The confinement leads to discrete
energy levels which depend on the film thickness. The total Hamiltonian
is given by: 
\begin{equation}
\mathcal{H}=\mathcal{H}_{0}+\mathcal{H}_{1}.\label{eq:H}
\end{equation}
The first term describes a film of constant thickness $d$, without
roughness represented by: 
\begin{equation}
\mathcal{H}_{0}=\frac{|\vec{p}|^{2}}{2m}+u_{d}\left(z\right)=\sum_{nq}\varepsilon_{nq}a_{nq\sigma}^{+}a_{nq\sigma},\label{eq:H0}
\end{equation}
where $m$ is the mass of electron and $\vec{p}=i\hbar\vec{\nabla}$
is the momentum operator, $u_{d}\left(z\right)$ is the confined potential
of a particle in a box. The right term in Eq.(\ref{eq:H0}) is the
kinetic energy of conduction electrons with energy $\varepsilon_{nq}=\hbar^{2}q^{2}/2m^{*}-\varepsilon_{F}$
measured from Fermi level $\varepsilon_{F}$, $q=\left(q_{x},q_{y}\right)$
is in plane wave vector and $m^{*}$ is effective mass, where $\frac{1}{m^{*}}=\frac{1}{m}\left(\frac{k_{n}^{2}}{q^{2}}+1\right)$,
$k_{n}=n\pi/d$, and $n$ denotes transverse mode and the operator
$a_{nq\sigma}^{+}\left(a_{nq\sigma}\right)$ creates (annihilates)
an electron with spin $\sigma$ at $n$th level. The second term in
Eq.(\ref{eq:H}) $\mathcal{H}_{1}$ becomes, 
\begin{align}
\mathcal{H}_{1}= & \sum_{nn'}\sum_{qq'}\sum_{\sigma\sigma'}\left\langle n'q'\sigma'\left|U\right|nq\sigma\right\rangle a_{n'q'\sigma'}^{+}a_{nq\sigma},\label{eq:H1}
\end{align}
where $\left\langle ...\right\rangle $ denotes the expectation value
over the state $\left|nq\sigma\right\rangle =\sqrt{2/\mathcal{A}d}\sin\left(k_{n}z\right)\exp\left(iq.\rho\right)\left|\sigma\right\rangle $,
$\mathcal{A}$ is the lateral area, $\left|\sigma\right\rangle $
is the eigenspin state with $\hat{\sigma}\left|\sigma\right\rangle =\sigma\left|\sigma\right\rangle $
and $\hat{\sigma}=\left(\sigma_{x},\sigma_{y},\sigma_{z}\right)$
is the Pauli spin operator. The total impurity potential $U$ is the
sum of the bulk impurity $u_{I}$, surface roughness scattering $u_{R}$
and the spin-orbit interaction due to the surface scattering $u_{R}^{SO}$,
$U=u_{I}+u_{R}+u_{R}^{SO}.$ Summarizes, Eq.(\ref{eq:H1}) is the
scattering of conduction electrons between different transverse modes
of momentum and spin states. Now we concentrate on scattering mechanism
treated as perturbation and introduce three terms as original reasons
for it. First of all, the bulk impurity scattering can be addressed
with a short range impurity potential with concentration $n_{imp}$,
\begin{equation}
u_{I}(r)=\frac{u_{imp}}{k_{F}^{3}}\sum_{i}\delta_{D}(r-r_{i}),\label{eq:ui}
\end{equation}
where $\delta_{D}$ is Dirac delta function. The magnitude of the
potential is given by $u_{imp}$ , $r_{i}=\left(\rho_{i},z_{i}\right)$
stands for the position of $i$th impurity and $k_{F}$ is the Fermi
wave vector. Next, we consider a rough surface, by a dilation operator
with $\lambda_{\rho}=\ln\frac{d}{d(\rho)}$ the surface roughness
is converted into effective scattering potential \cite{tevsanovic1986quantum,trivedi1988quantum}.

\begin{equation}
u_{R}=\lambda_{\rho}\left(2u_{d}+z\partial zu_{d}\right).\label{eq:ur}
\end{equation}
We assume that the volume of the original film with rough surface
remains unchanged after dilation transformation so the ensemble average
over roughness profile equal to $\left\langle d\left(\rho\right)\right\rangle =d$,
and hence $\left\langle \lambda_{\rho}\right\rangle =0.$ Furthermore,
we consider the surface roughness is correlated and to be isotropic,
$C\left(\rho=\left|\rho'-\rho''\right|\right)=\left\langle \lambda_{\rho'}\lambda_{\rho''}\right\rangle $.
In addition, the gradient of the effective scattering potential leads
to a SOC \cite{zhou2015spin}, as 
\begin{equation}
u_{R}^{SO}=\frac{\eta}{\hbar}\sigma\cdot\left(\vec{p}\times\nabla u_{R}\right),\label{eq:urso}
\end{equation}
where $\eta$ is the SOC parameter for the surface scattering and
$\hbar$ is the reduced Planck's constant.

Considering Eqs. (\ref{eq:ui}), (\ref{eq:ur}) and (\ref{eq:urso}),
the second term in Eq. (\ref{eq:H}) becomes 
\begin{alignat}{1}
\left\langle n'q'\sigma'\left|U\right|nq\sigma\right\rangle  & =u_{q'q}\delta_{\sigma\sigma'}\delta_{nn'}+u_{R_{n'q'\,nq}}\delta_{\sigma\sigma'}\delta_{nn'}\nonumber \\
 & +i\eta u_{R_{n'q'\,nq}}[\sigma_{\sigma\sigma'}\cdot(q'\times q)]\delta_{nn'},
\end{alignat}
where $u_{q'q}=\left\langle q'\left|u_{I}\right|q\right\rangle =\left(u_{imp}/V\right)\sum_{i}\exp[i\left(q-q'\right)\cdot r_{i}]$,
$V$ is the volume of the film. $u_{R_{n'q'\,nq}}=\left\langle n'q'\left|u_{R}\right|nq\right\rangle =\lambda_{q-q'}\varepsilon_{0}nn'$
with $\varepsilon_{0}=\hbar^{2}\pi^{2}/2md^{2}$ and $\lambda_{q-q'}$
is the Fourier component of $\lambda_{\rho}$ of the wave vector $q-q'$.

\section*{III. Conductivities}

\subsection*{A. Longitudinal conductivity}

Based on Lippman-Schwinger formalism we can calculate the transition
probability, $P$ from state $\left|nq\sigma\right\rangle $ to $\left|n'q'\sigma'\right\rangle $,
\begin{equation}
P_{nq\sigma}^{n'q'\sigma'}=\frac{2\pi}{\hbar}\left|\left\langle n'q'\sigma'\left|T\right|nq\sigma\right\rangle \right|^{2}\delta_{D}\left(\varepsilon_{nq\sigma}-\varepsilon_{n'q'\sigma'}\right),\label{eq:P}
\end{equation}
where $T=U+U(E-\mathcal{H})^{-1}U$ is the scattering matrix, whose
matrix elements are calculated up to the second-order Born approximation.
The relaxation time allows us to compute the conductivities. The relaxation
rate for each channel at Fermi energy is obtained from the transition
probability shown in Eq.$(\ref{eq:P})$, after ensemble averaging
over surface profiles and employing Matthiessen’s rule, we can find
relaxation times. The total scattering rate is, 
\begin{equation}
\frac{1}{\tau_{n}}=\sum_{n'q'\sigma'}P_{nq\sigma}^{n'q'\sigma'}=\frac{1}{\tau_{0}}+\frac{1}{\tau_{n}^{'}},
\end{equation}
where 
\begin{equation}
\frac{1}{\tau_{0}}=\frac{u_{imp}^{2}}{\varepsilon_{F}\hbar}\,\frac{n_{imp}}{2\pi k_{F}^{3}n_{c}}\left(1+2n_{c}\right),
\end{equation}
and

\begin{equation}
\frac{1}{\tau_{n}^{'}}=\frac{2}{3}\frac{\varepsilon_{F}}{\hbar}\,\frac{\nicefrac{{\scriptstyle {\textstyle {\scriptscriptstyle {\scriptstyle \sum_{n=1}^{n_{c}}}}}}n^{2}}{n_{c}^{3}}}{n_{c}\left({\scriptstyle {\scriptstyle \sum_{n=1}^{n_{c}}}}\frac{1}{n^{2}}-1\right)}\left\langle \left|S\left(q\right)\right|^{2}\right\rangle ,\label{eq:tua prime}
\end{equation}
\begin{figure}
\includegraphics{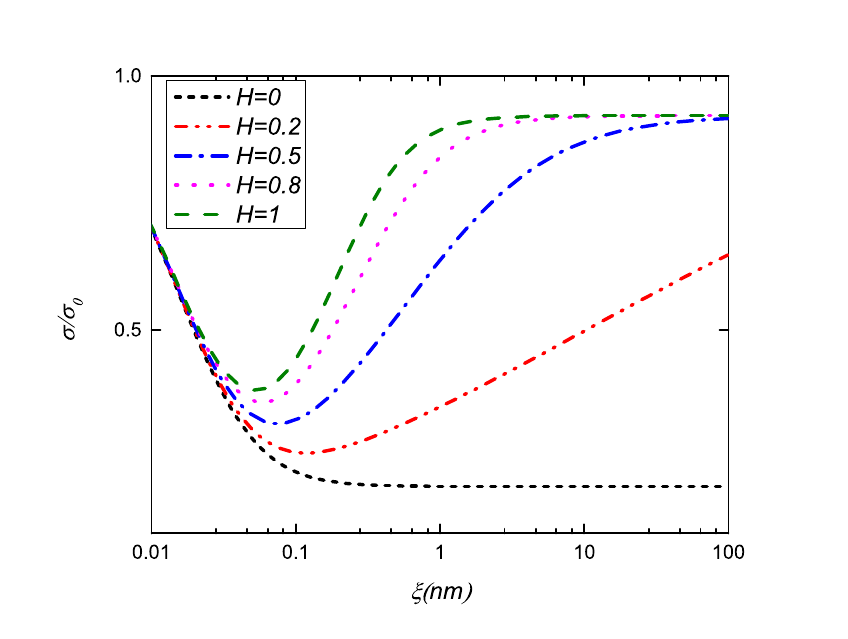}\caption{Longitudinal conductivity $\sigma$,\label{sigma-correlation} for
$Cu$ film vs. correlation length $\xi$ for $\varepsilon_{F}=7eV$,
$\tau_{0}=24fs$, $q=k_{F}=1.36\,1/\mathring{A}$, $n_{c}=10$, $\delta=5a_{0}$,
$a_{0}=3.61\mathring{A}$, $\sigma_{0}=5.88\times10^{7}s/m$, and
$H$, as indicated.}
\end{figure}
where $\tau_{0}^{-1}$ is the bulk impurity scattering rate and $\tau_{n}^{'-1}$
is the channel (n) dependent surface scattering rate, $\left\langle \left|S\left(q\right)\right|^{2}\right\rangle $
is the Fourier transform $C(\rho)$ which is called power spectrum
and the total number of transverse channels is $n_{c}=k_{F}d/\pi$
and $n\leq n_{c}$. In Eq.\ref{eq:tua prime} we show that $\tau_{n}^{'-1}$
is dependent to power spectrum or height-height correlation function
therefore by changing of $C(\rho)$ we can make a difference in $\tau_{n}^{'-1}$
and optimize SHC and SHA. 

\begin{figure}
\includegraphics{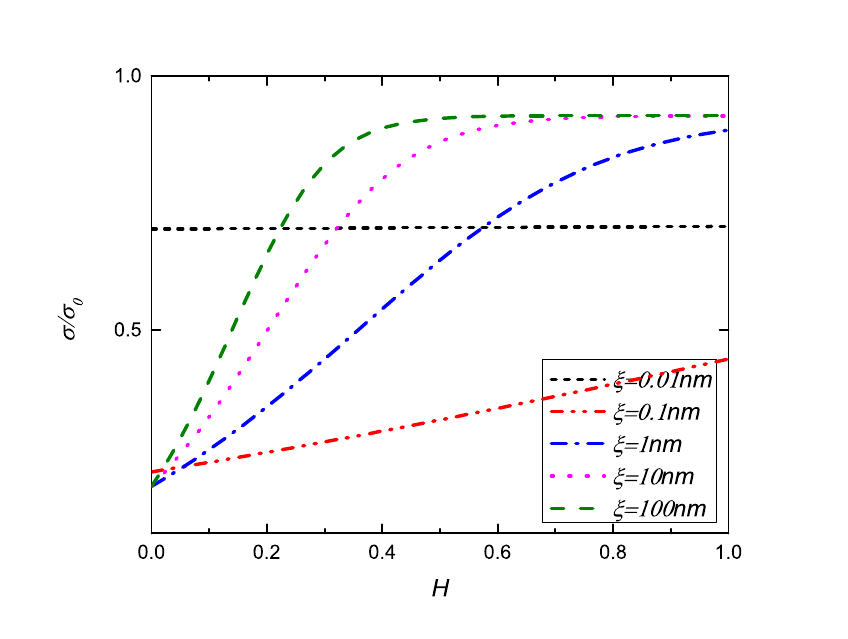}\caption{$\sigma$ a\label{sigma-h}s a function of the Hurst exponent $H$
with same parameters used in Fig.\ref{sigma-correlation} and indicated
values of correlation length $\xi$.}
\end{figure}

Employing the aforementioned formalism results in longitudinal conductivity
for any correlated surface roughness by \cite{tevsanovic1986quantum,trivedi1988quantum}

\begin{equation}
\sigma=\frac{3\sigma_{0}}{2n_{c}}\sum_{n=1}^{nc}\frac{\tau_{n}}{\tau_{0}}\left(1-\frac{n^{2}}{n_{c}^{2}}\right),\label{eq:sigma}
\end{equation}
where $\sigma_{0}=k_{F}^{3}e^{2}\tau_{0}/3\pi^{2}m^{*}$ is the bulk
Drude conductivity. In Fig.\ref{sigma-correlation} we present the
room temperature longitudinal conductivity versus correlation length
$\xi$ for $Cu$ thin film, for several values of the roughness exponent
$H$. There is a minimum in the film conductivity as a function of
the correlation length, which occurs approximately at $\xi=0.1nm\sim\mathcal{O}\left(a_{o}\right)$,
where $a_{0}$ is the lattice constant. For $\xi>>0.1nm$ the conductivity
increases with increasing $H$ or increasing $\xi$ (smoother surface).
For small values of $\xi<<0.1nm$ the situation is reversed (in this
case the correlation length is so smaller than the lattice constant,
therefore we can suppose that the surface is uncorrelated $\xi\sim0$).
This is due to the fact that this kind of roughness does not scatter
electrons when their wavelength is much longer than the correlation
length $\xi$. For large values of $\xi$ the behavior is more complex,
i.e., the conductivity reaches to a maximum with strong dependency
on $H$. Fig.\ref{sigma-h} shows the dependence of the film conductivity
on the roughness exponent $H$ shown for several values of the correlation
length $\xi$. For $\xi=10,100\,nm$, the conductivity first increases
with increasing $H$, with a further increase of the roughness exponent
$H$. For the correlation lengths longer than $a_{0}$ the conductivity
increases with increasing $\xi$ at a much faster rate for large $H\,(H\sim1)$.
Additionally, with larger correlation length this maximum point shifts
to the smaller roughness exponent. For the extreme limit, as can be
seen from the curve that correspond to the logarithmic roughness$(H=0)$
in Fig.$\ref{sigma-h}$, the conductivity increases extremely slowly
with increasing correlation length. Thus, the smoothing effect at
large length scale is strongly influenced by the roughness exponent.
The result agrees with those obtained in Ref \onlinecite{m}. 

\subsection*{B. Spin Hall conductivity}

SHE originates from three mechanisms, the SS, SJ and intrinsic \cite{maekawa2012spin},
\begin{equation}
\sigma^{SH}=\alpha_{ss}^{SH}\sigma+\alpha_{sj}^{SH}\sigma+\sigma_{int}^{SH},\label{eq:sigmash}
\end{equation}
here $\alpha_{ss}^{SH}$, is the SHA from skew mechanism. If roughness
scattering potential distribution function is symmetrical, it means
that the third moment or skewness of scattering potential is zero
$\left\langle U^{3}\right\rangle =0$, then $\alpha_{ss}^{SH}=0$,
$\left\langle ...\right\rangle $ denotes the ensemble average over
roughness profiles. Therefore, if the distribution function of surface
roughness taken to be Gaussian, the SS does not have any contribution
in SHC induced by surface roughness. Contrarily, for that to be non-Gaussian,
the SS can contribute in SHC because of $\left\langle U^{3}\right\rangle \neq0$.
The intrinsic SHC is negligible, i.e. $\sigma_{int}^{SH}=0$ as we
consider the films with very weak bulk SOC such as $Cu$. For calculation
of $\alpha_{sj}^{SH}$, SHA due to SJ mechanism, we need to compute
the velocity of an electron $\mathcal{V}_{nq}^{\sigma}$ in the presence
of the spin-orbit potential $u_{R}^{SO}$. It can be found by calculating
the matrix element $\mathcal{V}_{nq}^{\sigma}=\left\langle nq^{+}\sigma\left|\mathcal{\hat{V}}\right|nq^{+}\sigma\right\rangle $
of the velocity operator 
\begin{equation}
\mathcal{\hat{V}}=\frac{1}{i\hbar}\left[r,\mathcal{H}\right]=\frac{p}{m}+\frac{\eta}{\hbar}\left(\sigma\times\nabla u_{R}\right),
\end{equation}
\begin{figure}
\begin{singlespace}
\includegraphics{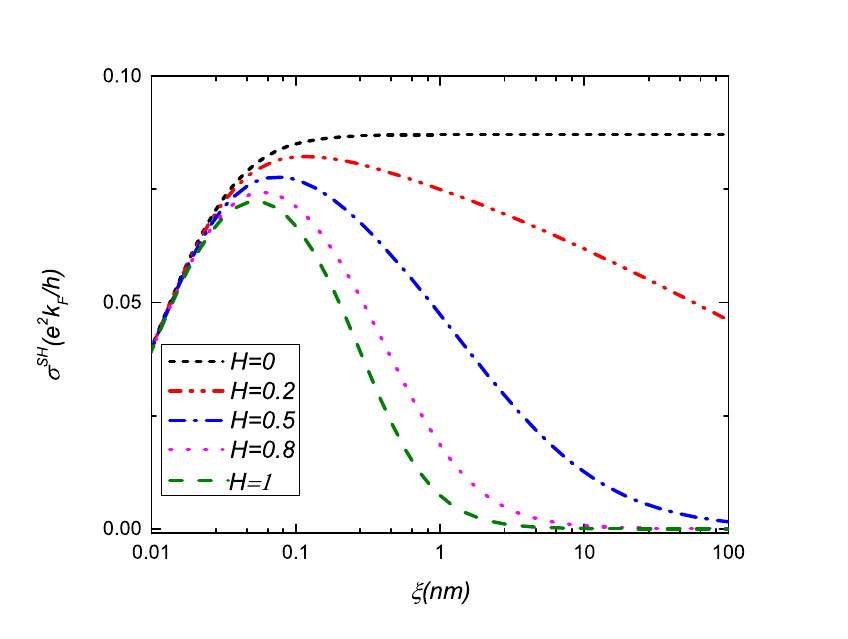}\caption{\label{sigmash-cor}SHC $\sigma^{SH\,}$vs. correlation length $\xi$
for $Cu$ film with same parameters used in Fig.\ref{sigma-correlation},
and $\bar{\eta}=0.5$ and indicated values of $H$.}
\end{singlespace}
\end{figure}
between the scattering state $\left|nq^{+}\sigma\right\rangle =\left|nq\sigma\right\rangle +\sum_{n'q'}u_{q'q}(\varepsilon_{nq}-\varepsilon_{n'q'}+i\epsilon)^{-1}\left|n'q'\sigma\right\rangle $
within the Born approximation, and becomes 
\begin{equation}
\mathcal{V}_{nq}^{\sigma}=v_{nq}+\omega_{nq}^{\sigma},\:\omega_{nq}^{\sigma}=\alpha_{sj}^{SH}\left(\sigma_{\sigma\sigma}\times v_{nq}\right),
\end{equation}
where $v_{nq}=\hbar q/m^{*}$ is the ordinary velocity, $\omega_{nq}^{\sigma}$
is the anomalous velocity, $\sigma_{\sigma\sigma}=\left\langle \sigma\left|\hat{\sigma}\right|\sigma\right\rangle $
is the polarization vector. SHA due to SJ is $\alpha_{sj}^{SH}=\frac{\hbar\eta}{2\varepsilon_{F}\tau'_{n}}$,
by substituting of $\alpha_{sj}^{SH}$ in Eq.$(\ref{eq:sigmash})$
the SHC becomes,

\begin{equation}
\sigma^{SH}=\frac{e^{2}k_{F}}{h}\frac{\bar{\eta}}{n_{c}\pi}\sum_{n=1}^{n_{c}}\frac{\tau_{n}}{\tau_{n}^{'}}\left(1-\frac{n^{2}}{n_{c}^{2}}\right),\label{eq:sigmash1}
\end{equation}
where $\bar{\eta}=k_{F}^{2}\eta$ and $h$ is the Planck's constant.

Therefor, we show that if distribution function of surface roughness
to be Gaussian the SHE is only from the SJ contribution and for that
to be non-Gaussian, both SJ and SS are present the SJ has the main
contribution to the SHC.

\begin{figure}
\includegraphics{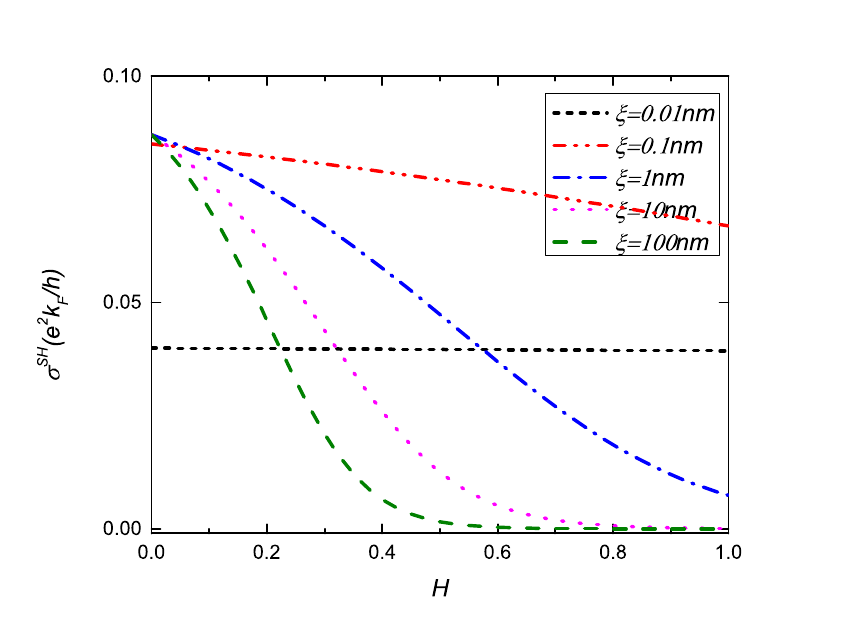}\caption{\label{sigmash-h}SHC $\sigma^{SH}$ as a function of Hurst exponent
$H$ for $Cu$ film with $\bar{\eta}=0.5$ and same parameters used
in Fig.\ref{sigma-correlation}, and indicated values of $\xi$.}
\end{figure}

\section*{IV. Roughness surface model}

In this section, we investigate the influence of uncorrelated and
self-affine surface profiles on SHC.

\subsection*{A. Uncorrelated surface profile}

In this case, the surface roughness is uncorrelated and the correlation
function and the its Fourier transform becomes, 
\begin{equation}
C\left(\rho\right)=\varLambda a_{0}^{2}\delta_{D}\left(\rho'-\rho''\right),
\end{equation}
and 
\begin{equation}
\left\langle \left|S(q)\right|^{2}\right\rangle =aa_{0}^{2}\varLambda,\label{eq:whitenoise}
\end{equation}
where $\varLambda=\left(\frac{\delta}{d}\right)^{2}$ is the dimensionless
parameter, $\delta$ is variance of height fluctuations. The normalization
condition $\int_{0<q<q_{c}}\left\langle \left|S(q)\right|^{2}\right\rangle d^{2}q=\left(\frac{2\pi}{a_{0}}\right)^{2}\varLambda$
yields the parameter $a$. This parameter is used to satisfy of the
normalization condition for Fourier transform of correlation function.
Here, $q_{c}=\pi/a_{0}$, is the upper cutoff in the Fourier space
where $a_{0}\sim k_{F}^{-1}$ is the lattice constant. Surface scattering
relaxation time $\tau'_{n}$ obtained by substituting Eq.$(\ref{eq:whitenoise})$
in Eq.$(\ref{eq:tua prime})$. By substituting $\tau_{n}'$ in $Eq.(\ref{eq:sigma})$
and $\left(\ref{eq:sigmash1}\right)$ the longitudinal conductivity
and SHC can be obtained. The results agree with those found in Ref.
\onlinecite{zhou2015spin}.

\subsection*{B. Self-affine fractal surface profile}

For a typical self-affine surface with power-law correlation function
$C(\rho)$ is characterized by a correlation length $\xi$ \cite{m,makse1996method},
\begin{equation}
C\left(\rho\right)=\varLambda\left(1-\left(\frac{\rho}{\xi}\right)^{2H}\right),
\end{equation}
the roughness exponent$0\leq H\leq1$ is a measure of the degree of
surface irregularity. Small values of $H$ characterize jagged or
irregular surfaces at short length scales $(\rho<<\xi)$, where the
correlation function shows power-law behavior, while large values
of $H$ correspond to smoother height-height fluctuations. The Fourier
transform of correlation for self-affine fractals has the scaling
behavior if $q\xi>>1$ and for white noise profile that occurs if
$q\xi<<1$. We use k-correlation model for description of self-affine
surface, 

\begin{figure}
\includegraphics{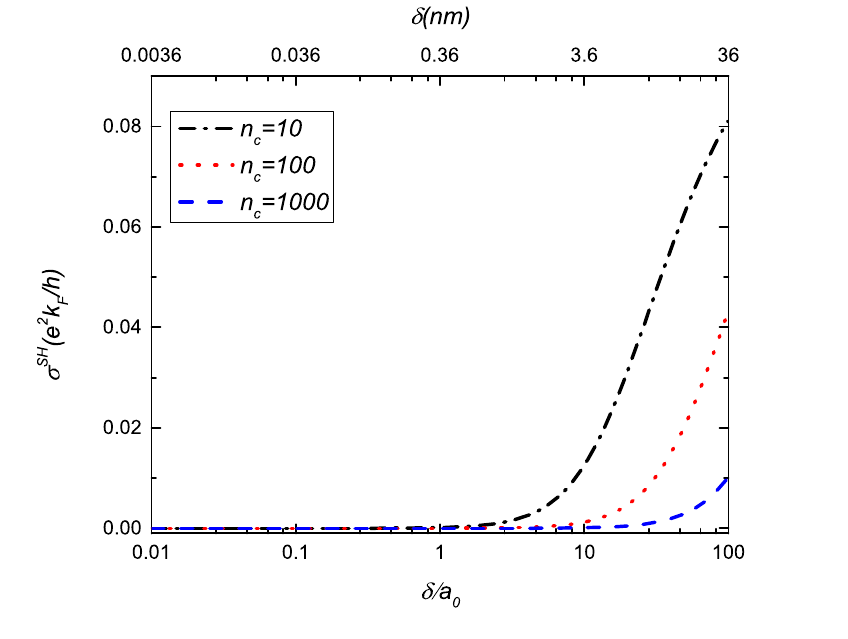}\caption{\label{sigmsh-delta}SHC $\sigma^{SH}$ for $Cu$ thin film as a function
of $\delta$ for three values of film thickness $n_{c}=10,\,100,\,1000$
with $\varepsilon_{F}=7eV$, $q\sim k_{F}=1.36$$\:\unit{1/\mathring{A}}$
,$a_{0}=3.61\:\unit{\mathring{A}}$,$\sigma_{0}=5.88\times10^{7}\:\unit{\unit{\unitfrac{S}{m}}},\tau_{0}=24\:\unit[f]{s},$$\bar{\eta}=0.5$,
$\xi=10nm$ and $H=0.5$.}
\end{figure}

\begin{equation}
\left\langle \left|S(q)\right|^{2}\right\rangle =\frac{2\pi}{a_{0}^{2}}\frac{\xi^{2}\varLambda}{\left(1+aq^{2}\xi^{2}\right)^{1+H}},\label{eq:self affine}
\end{equation}
the normalization condition yields the parameter $a$, 
\begin{equation}
a=\begin{cases}
(1/2H)\left[1-\left(1+aq_{c}^{2}\xi^{2}\right)^{-H}\right] & 0<H\leq1\\
\left(1/2\right)\ln\left(1+aq_{c}^{2}\xi^{2}\right) & H=0
\end{cases}.
\end{equation}
By substituting Eq.$(\ref{eq:self affine})$ into Eq.$(\ref{eq:tua prime})$
and by substituting $\tau_{n}'$ into Eq.$\left(\ref{eq:sigma}\right)$
and $\left(\ref{eq:sigmash1}\right)$ we are able to investigate the
effect of fractal surface in SHC and SHA.

In Fig.\ref{sigmash-cor} we present the SHC $\sigma^{SH}$, versus
correlation length $\xi$ for several values of roughness exponent
$H$. A characteristic feature seen in Fig.\ref{sigmash-cor} is the
presence of a maximum in the $\sigma^{SH}$ as function of the roughness
exponent $H$, which occurs approximately at $\xi=0.1nm\sim\mathcal{O}\left(a_{o}\right)$.
For large values of $\xi$, the $\sigma^{SH}$ exhibits a normal behavior.
It decreases with increasing $\xi$ or increasing $H$ (surface smoothing)
and for small values of $\xi\ll0.1nm$ (in this case the correlation
length is so smaller than the lattice constant, therefore we can suppose
that the surface is uncorrelated $\xi\sim0.$) the situation is reversed.
This is due to the fact that this kind of roughness is in a range
$q\xi<<1$ . In this range, the power spectrum in power-law model
is white noise and does not have scaling behavior. Also in large scale,
in the logarithmic roughness $(H=0)$, the SHC increases extremely
slowly with increasing correlation length. In Fig.\ref{sigmash-h}
for several values of correlation length we present the SHC as a function
of roughness 
\begin{figure}
\includegraphics{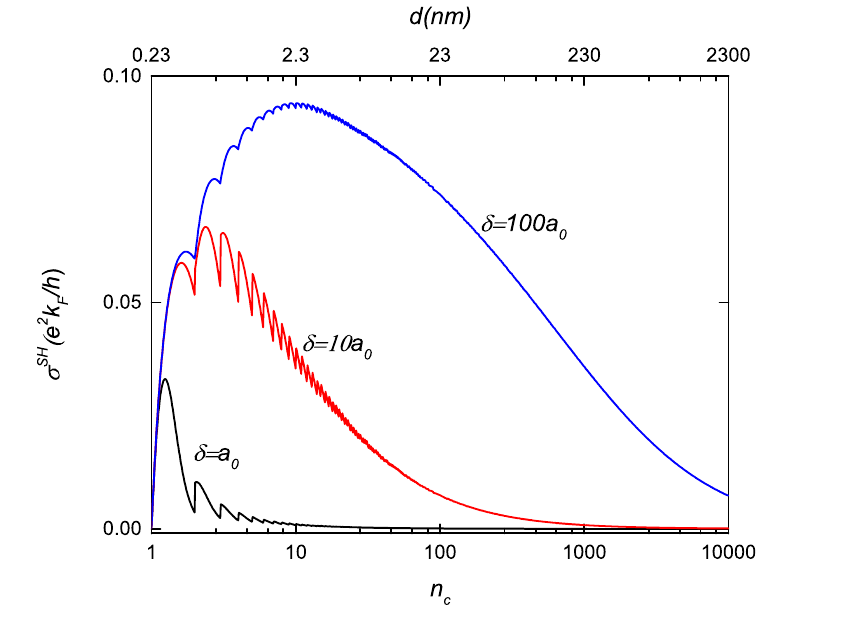}\caption{\label{sigmash-nc}Spin Hall conductivity $\sigma^{SH}$ for $Cu$
thin film as a function of $n_{c}$ with $\xi=1nm$ and same parameters
as in Fig.\ref{sigmsh-delta}, and indicated value of $\delta$. }
\end{figure}
exponent $H$ for $Cu$ thin film. Note that $\sigma^{SH}$ is larger
for a rougher surface with smaller correlation length.

The dependence of SHC on the $\delta$ and $n_{c}$ for $Cu$ thin
film with self-affine roughness profile is shown in Fig.\ref{sigmsh-delta}
and Fig.\ref{sigmash-nc}. For $\delta/nc<a_{0}$ thin films the conductivity
is independent on $\delta$ and for $\delta/nc>a_{0}$ the SHC increases
with increasing ratio $\delta/n_{c}$, in agreement with those in
Ref. \onlinecite{zhou2015spin}. Also in Fig.\ref{sigmash-nc} the
thickness dependence of SHC is plotted and is found to show oscillatory
behavior dependent upon $\delta$ in $nc\sim5-30$ because of quantum
size effect \cite{trivedi1988quantum}. In addition, the conductivity
has a maximum in the dependence on $\delta$ then it decreases with
increasing $n_{c}$.

\begin{figure}
\includegraphics{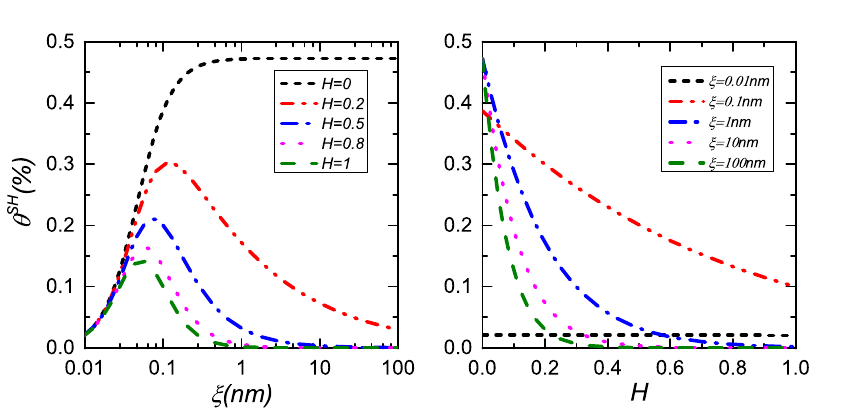}\caption{\label{SHA} The SHA $\theta(\%)$ for $Cu$ film with $n_{c}=100$,
$\delta=5a_{0}$, $\bar{\eta}=0.5$ and the same parameter used in
Fig.\ref{sigma-correlation}. Left panel shows the SHA as a function
of correlation length $\xi$ and right panel shows the SHA as a function
of Hurst exponent $H$.}
\end{figure}

Thus in the thinner and rougher film, $\sigma^{SH}$ is larger and
at large scale of the correlation length $\left(d<<\xi\right)$ gets
strongly influenced by the roughness exponent. The SHC and SHA can
increase by one order of magnitude at large scale of the $\xi$ when
the roughness exponent varies from $H=1$ to $H=0$. As shown in Fig.\ref{SHA}
the SHA in large scale can be enhanced by $(i)$ decreasing correlation
length $\xi$, $(ii)$ decreasing roughness exponent $H$ and in small
scale the situation is reversed. Based on experiments, the general
trends for the thin films is that the SHC increases with film roughness
\cite{akyol2016effect,an2016spin} which seems to agree with our aforementioned
argument. In general, one tends to use the interface width (root-mean
square roughness)$\delta$ to measure how rough the surface is: if
$\delta$ is large, then the surface is rougher. However, we have
demonstrated that if the surface is self-affine the SHC and SHA depends
not only on $\delta$ but also on the film thickness $d$, the lateral
correlation length $\xi$, and the roughness exponent $H$.

\section*{V. Conclusion}

In conclusion, we study the contribution of surface roughness and
the fractality effects on the SHE of metallic thin films with self-affine
correlation as a more realistic model. In the group of three surface
roughness parameters $(\delta,\xi,H)$, major interplay of the roughness
effect occurs for $H$ and $\xi$. The parameter $\delta$ has a minor
effect (especially in thicker films) since it appears in the form
a multiplication factor $(\sigma^{SH}\sim\delta^{-2})$. The roughness
exponent $H$ has a powerful impact on the SHC mainly for relatively
large correlation lengths that can increase the SHA for $Cu$ thin
film one order of magnitude. The SHA due to fractal roughness scattering
increases with reducing the correlation length and Hurst exponent
and film thickness in large scale whereas the situation is reversed
in small scale. Moreover, we found that if distribution function of
surface roughness to be Gaussian the SHE is only from the SJ contribution
and for that to be non-Gaussian, both SJ and SS are present. Our results
uncover additional contribution from statistical information at surface
of thin films which have important effects in the SHE.

\bibliographystyle{apsrev4-1}
\bibliography{7C__Users_roomate_Desktop_spintronic1_my_paper_1_final_1_after_prb_reject_88}

\end{document}